\documentclass[twocolumn,noshowpacs,oneside,prb,balancelastpage]{revtex4}%
\usepackage{amsfonts}
\usepackage{amsmath}
\usepackage{amssymb}
\usepackage{graphicx}%
\newtheorem{theorem}{Theorem}

\newtheorem{example}[theorem]{Example}

\begin{document}
\preprint{ }
\title[ ]{Electromagnetics from a quasistatic perspective}
\author{Jonas Larsson}
\affiliation{Deparment of Physics, Ume\aa \ University, SE-90187 Ume\aa , Sweden}
\pacs{PACS number}

\begin{abstract}
Quasistatics is introduced so that it fits smoothly into the standard textbook
presentation of electrodynamics. The usual path from statics to general
electrodynamics is rather short and surprisingly simple. A closer look reveals
however that it is not without confusing issues as has been illustrated by
many contributions to this Journal. Quasistatic theory is conceptually useful
by providing an intermediate level in between statics and the full set of
Maxwell's equations. Quasistatics is easier than general electrodynamics and
in some ways more similar to statics. It is however, in terms of interesting
physics and important applications, far richer than statics. Quasistatics is
much used in electromagnetic modeling, an activity that today is possible
on a PC and which also has great pedagogical potential. The use of
electromagnetic simulations in teaching gives additional support for the importance of quasistatics. This activity may also motivate some change of focus in the presentation of basic electrodynamics.\end{abstract}
\volumeyear{year}
\volumenumber{number}
\issuenumber{number}
\eid{identifier}
%\date[Date text]{date}
\received[Received text]{date}

\revised[Revised text]{date}

\accepted[Accepted text]{date}

%\published[Published text]{date}

\maketitle

\section{Introduction}

Applications of electrodynamics may be in the \textit{static},
\textit{quasistatic} or general \textit{high frequency} regime. Quasistatics is more or less neglected in most textbooks and the purpose of this paper is to present material that can fill this gap
in a course on the level of for example Griffith's
textbook.\cite{Griffiths} There are good reasons to do so including:

\begin{itemize}
\item[1.] The step from statics to general electrodynamics is huge in terms of
its physical content. Quasistatic models are useful by providing intermediate
levels in the theory. Thereby several confusing issues that result from the
very condensed standard derivation of Maxwell's equations may be avoided.
Questions concerning Coulomb's and Biot-Savart's laws in non-static situations
should be addressed.\cite{GriffithsHeald,FrenchTessman,Bartlett} So should
also the appearance of dynamic electric fields in regions where the magnetic
field seems to be absent (like outside a long solenoid or toroidal coil with a
time-varying current).\cite{French,Mills,Walstad,Templin,Protheroe} Another
confusing issue follows from the standard textbook motivation for the
displacement current where it is introduced in order to make the equations of
electrodynamics consistent with charge conservation. The physical impact of
this term is remarkable and includes in particular electromagnetic waves in
free space. But why should waves in \textit{free space without charges and
currents} follow from \textit{charge conservation}?\cite{Zapolsky}
Quasistatics is useful in the discussion of these and many other questions.

\item[2.] There are plenty of interesting phenomena within the quasistatic
regime. Using the full set of Maxwell's equations are many times unnecessarily
complicated since these equations can describe the most intricate
electromagnetic wave phenomena involving short time-scales or high frequency.
Such an analysis may be difficult and is not necessary in quasistatic
situations.\cite{HausMelcher}

\item[3.] Many real world applications involve the numerical solution of the
time-dependent Maxwell equations in three dimensions. However, quite often one
is interested in phenomena where some quasistatic analysis is sufficient. This
amounts to the replacement of an hyperbolic model with an elliptic (or
parabolic) one which can be solved in more economic
ways.\cite{NielsonLewis,HewittBoyd,GibbonsHewitt}

\item[4.] Basic quasistatic theory may be introduced in an elementary and
simple way suitable for basic courses in electrodynamics. Like static theory
also quasistatics may be given two equivalent formulations. The first is in
terms of force laws, including straightforward generalizations of the static
Coulomb and Biot-Savart laws. The second formulation is in terms of simple
approximations in Maxwell's differential equations.
\end{itemize}

This paper is organized as follows. Basic theory of quasistatics is presented
in Section II. Subsections A and B contain the two equivalent formulations of
quasistatics, one in terms of force laws and the other in terms of
approximations in Maxwell's differential equations. A salient feature of
quasistatics is instantaneous interaction at a distance. In subsection C
the corresponding $c\rightarrow\infty$ limit of general electrodynamics is
used to derive quasistatics. Subsections D, E and F include quasistatic
theory for the electromagnetic potentials, the fields from a moving point
charge and the quasistatic Poynting theorems. Alternatives to the standard textbook
derivation of Maxwell's equations are discussed in Section III from a
quasistatic perspective.

Sections IV and V concern some confusing issues that have been discussed in
many contributions to this Journal. In Section IV we consider the laws of
Coulomb and Biot-Savart within general time-dependent
theory.\cite{GriffithsHeald} Sufficient and necessary conditions for the exact
validity of these laws are given. Section V contains comments on the old
question \#6 of this Journal which concerns the fields outside a solenoid with a
time-varying current.\cite{French}\ \ 

Quasistatics is useful not only for providing improved understanding of basic
theory but it is also important for applications, in particular for numerical
simulations. In Section VI we discuss the possibility of using electromagnetic
simulations in basic courses and exemplify with quasistatic equations for eddy
currents. A couple of PDE-solvers which have been used for electromagnetic
modeling on a PC is mentioned. A summary is given in Section VII.

\section{Quasistatics}

\textit{Quasistatics} within electrodynamics refers to a regime where "the
system is small compared with the electromagnetic wavelength associated with
the dominant time-scale of the problem".\cite{Jacksons bok} \ The fields are
then propagated instantaneously so we are dealing with a kind of
$c\rightarrow\infty$ limit. We consider in this paper three major
\textit{quasistatic models}. These are EQS
(electroquasistatics)\cite{HausMelcher}, MQS
(magnetoquasistatics)\cite{HausMelcher} and the Darwin
model\cite{NielsonLewis,HewittBoyd,GibbonsHewitt}. EQS includes capacitive but
not inductive effects, MQS includes inductive but not capacitive effects while
the Darwin model includes both capacitive and inductive effects. The
Biot-Savart law is valid in all three models while the Coulomb law is valid
only in EQS. In MQS and Darwin there is also, besides the Coulomb field, an
additional contribution to the electric field due to magnetic induction. The
source of this electric field is thus $\partial\mathbf{B}/\partial t$ in
Faraday's law.\cite{dB/dt}

\subsection{From the laws of Coulomb, Biot-Savart and Faraday to the EQS, MQS
and Darwin models.}

Dynamical systems that proceed from one state to another as though they are
static (at each fixed time) are commonly said to be \textit{quasistatic}. For
electromagnetism the static theory builds on the Coulomb and Biot-Savart laws
together with the static continuity equation. Quasistatics would then simply
be obtained by the allowance for time-dependence in the first two otherwise
unchanged laws%

\begin{equation}
\mathbf{E}\left(  \mathbf{r},t\right)  =\dfrac{1}{4\pi\varepsilon_{0}}%
\iiint\dfrac{\rho\left(  \mathbf{r^{\prime}},t\right)  \mathbf{\hat{R}}}%
{R^{2}}d\tau^{\prime} \label{coul}%
\end{equation}

\begin{equation}
\mathbf{B}\left(  \mathbf{r},t\right)  =\dfrac{\mu_{0}}{4\pi}\iiint
\dfrac{\mathbf{J}\left(  \mathbf{r^{\prime}},t\right)  \times\mathbf{\hat{R}}%
}{R^{2}}d\tau^{\prime} \label{biot}%
\end{equation}
where we use the notations\cite{scriptR}
\begin{equation}
\mathbf{R}=\mathbf{r}-\mathbf{r}^{\prime},\ R=\left\vert \mathbf{R}\right\vert
,\ \ \mathbf{\hat{R}=R}/R \label{Rdef}%
\end{equation}
It would however appear strange to keep the static continuity equation
unchanged in this time-dependent situation so we replace it with the usual
continuity equation\cite{QS0}%

\begin{equation}
\dfrac{\partial\rho}{\partial t}+\triangledown\cdot\mathbf{J}=0, \label{cont}%
\end{equation}
The three equations (\ref{coul}), (\ref{biot}) and (\ref{cont}) constitute the
EQS model that includes capacitive but not inductive
effects. Charge may be accumulated in this model but this require work and
energy may as usual be associated with the electric field. Magnetic energy is
however outside the scope of EQS because there is no magnetically induced
electric field and accordingly no back emf. No work is then required to create a magnetic field by starting
an electric current.

We now like to include electromagnetic induction. Then also $\partial
\mathbf{B}/\partial t$ acts as a source of electric fields and
the total electric field becomes the sum of two parts
\begin{equation}
\mathbf{E=E}_{C}\mathbf{+E}_{F} \label{EC+EF}%
\end{equation}
The first term is the Coulomb electric field%

\begin{equation}
\mathbf{E}_{C}\left(  \mathbf{r},t\right)  =\dfrac{1}{4\pi\varepsilon_{0}%
}\iiint\dfrac{\rho\left(  \mathbf{r^{\prime}},t\right)  \mathbf{\hat{R}}%
}{R^{2}}d\tau^{\prime} \label{EC}%
\end{equation}
and the second term is the Faraday electric field which may be defined by a Biot-Savart like integral expression as (cf. equation \ref{biot}) \cite{FaradayIntRemark}%
\begin{equation}
\mathbf{E}_{F}\left(  \mathbf{r},t\right)  =-\dfrac{1}{4\pi}\iiint
\frac{\partial\mathbf{B}\left(  \mathbf{r^{\prime}},t\right)  }{\partial
t}\times\dfrac{\mathbf{\hat{R}}}{R^{2}}d\tau^{\prime} \label{EF}%
\end{equation}
We note that $\mathbf{E}_{F}$ solves the equations%
\begin{equation}
\triangledown\times\mathbf{E}_{F}=-\dfrac{\partial\mathbf{B}}{\partial
t},\text{ \ \ \ }\triangledown\cdot\mathbf{E}_{F}=0\text{\ } \label{EFdiff}%
\end{equation}
while the corresponding equations for $\mathbf{E}_{C}$ is%
\begin{equation}
\triangledown\times\mathbf{E}_{C}=0,\ \ \ \triangledown\cdot\mathbf{E}%
_{C}=\frac{\rho}{\varepsilon_{0}} \label{ECdiff}%
\end{equation}
The expressions (\ref{EC}) and (\ref{EF}) are the unique solutions of
(\ref{EFdiff}) and (\ref{ECdiff}) provided appropriate boundary conditions at
infinity are used.\cite{Infinity}

The Darwin model may be defined by the equations (\ref{biot}) and
(\ref{cont})-(\ref{EF}). This is a quasistatic model that includes both
capacitive and inductive phenomena. Note that for given current and charge
densities we directly obtain the electromagnetic fields in terms of integrals
without the appearance of any time-retardation. Instead the integral
expressions in the laws of Coulomb and Biot-Savart play an important role in
this dynamical model.

The MQS model is obtained from Darwin if the usual continuity equation is
replaced by the static one%

\begin{equation}
\triangledown\cdot\mathbf{J}=0 \label{contstat}%
\end{equation}
Thus the MQS model may be defined by the equations (\ref{biot}),
(\ref{EC+EF}), (\ref{EC}), (\ref{EF}) and (\ref{contstat}). It is different
from both EQS and Darwin by including inductive but not capacitive
effects. A confusing feature of MQS is that the very fundamental continuity
equation may be violated by MQS-solutions. Only stationary currents are
allowed in MQS and these cannot explain changes in the charge density, thus
one cannot interpret the currents in MQS in terms of charge
transportation.\ Of course, only sufficiently good approximations of solutions
to the Maxwell equations are interesting in the real world so the continuity
equation must still be almost true in some sense. Amp\'{e}re's law, which
imply equation (\ref{contstat}), is valid in MQS but not in EQS or Darwin (see
the next subsection). Thus Amp\'{e}re's law is not always valid in quasistatics.
Griffiths and Heald remark that\ "The application of Amp\'{e}re's law in
quasistatic situations can be an extremely delicate
matter".\cite{GriffithsHeald}

The textbook of Haus and Melcher\cite{HausMelcher} builds up an understanding
of electrodynamics by using both EQS and MQS. This is of particular
significance in the relation between electromagnetic field theory and circuit
theory. Then EQS involves capacitance features and MQS inductance features.
For systems involving capacitance and inductance both models are needed. For
such applications it is crucial that capacitive and inductive
aspects are not both important in the same spatial place.\cite{CondEQSMQS} The use of two
complementary quasistatic models in the same physical system is clearly a
complicating feature if we like to model the whole system numerically. We
would then have to divide the whole spatial region into EQS and MQS subregions
with appropriate continuity conditions at the interfaces. A better alternative
may be to use the Darwin model which embrace all the physics contained in EQS
and MQS, still being quasistatic.

\subsection{From Maxwell's equations to EQS, MQS and Darwin.}

Let us now consider the formulation of quasistatics in terms of differential equations. The starting point is general electrodynamics with the Maxwell equations:

\begin{equation}
\triangledown\cdot\mathbf{E}=\left(  1/\varepsilon_{0}\right)  \rho
\label{gaussE}%
\end{equation}

\begin{equation}
\triangledown\cdot\mathbf{B}=0 \label{gaussB}%
\end{equation}

\begin{equation}
\triangledown\times\mathbf{E}=-\dfrac{\partial\mathbf{B}}{\partial t}
\label{faraday}%
\end{equation}

\begin{equation}
\triangledown\times\mathbf{B}=\mu_{0}\left(  \mathbf{J}+\varepsilon_{0}%
\dfrac{\partial\mathbf{E}}{\partial t}\right)  \label{ampmax}%
\end{equation}
These equations are complete and general as they stand. However, in the
presence of polarizable/magnetizable media it is in practice very convenient
to \textit{write}\ them in a different way by introducing the D- and the
H-fields. Then only the free charge and current densities appear explicitly in
the equations. For formal simplicity just E- and B-fields will be used in the
present paper but it is straightforward to modify the expressions so that a
formulation with D- and H-fields is obtained.

Let us now formulate the quasistatic models EQS, MQS and Darwin as approximations of
Maxwell's equations (\ref{gaussE})-(\ref{ampmax}). The relations to the
integral formulas in subsection A above will also be considered. The EQS model
is obtained from the Maxwell equations simply by neglecting $\partial
\mathbf{B}/\partial t$ in Faraday's law  while MQS is obtained by instead neglecting  $\partial\mathbf{E}/\partial t$ in the Amp\'{e}re-Maxwell law. Thus in EQS we have instead of (\ref{faraday}) the equation

\begin{equation}
\triangledown\times\mathbf{E}=0 \label{rotE}%
\end{equation}
and in MQS the equation (\ref{ampmax}) is replaced by the usual Amp\'{e}re
law\cite{Amperes lag def}

\begin{equation}
\triangledown\times\mathbf{B}=\mu_{0}\mathbf{J} \label{amp}%
\end{equation}

To get the Darwin model we do not neglect all of $\partial\mathbf{E}/\partial
t$ but keep the Coulomb part of the E-field
(defined by equation (\ref{EC}) above) and replaces the Amp\'{e}re-Maxwell law
with the Amp\'{e}re-Darwin equation \cite{ampere-darwin}%

\begin{equation}
\triangledown\times\mathbf{B}=\mu_{0}\left(  \mathbf{J}+\varepsilon_{0}%
\dfrac{\partial\mathbf{E}_{C}}{\partial t}\right)  \label{ampdarw}%
\end{equation}

Note an important difference between these two last equations. The Amp\'{e}re law
(\ref{amp}) implies equation (\ref{contstat}) and may violate charge
conservation in dynamic situations. The Amp\'{e}re-Darwin equation, like the
Amp\'{e}re-Maxwell equation, is consistent with the continuity equation (\ref{cont}).

\bigskip\begin{table}[ptb]
\caption{This table give two equivalent definitions of each of the quasistatic
models EQS, MQS and Darwin}%
\label{tab:table1}%
\begin{ruledtabular}
\begin{tabular}{lcr}
\textbf{Model} & \textbf{Def.A, eq.\#} & \textbf{Def.B, eq.\#}\\\hline
EQS & (\ref{coul}),(\ref{biot}),(\ref{cont}) & (\ref{gaussE}),(\ref{gaussB}%
),(\ref{ampmax}),(\ref{rotE})\\\hline
MQS & (\ref{biot}),(\ref{EC+EF}),(\ref{EC}),(\ref{EF}),(\ref{contstat}) &
(\ref{gaussE}),(\ref{gaussB}),(\ref{faraday}),(\ref{amp})\\\hline
Darwin & (\ref{biot}),(\ref{cont}),(\ref{EC+EF}),(\ref{EC}),(\ref{EF}) &
(\ref{gaussE}),(\ref{gaussB}),(\ref{faraday}),(\ref{ampdarw}), (\ref{ECdiff})\\
\end{tabular}
\end{ruledtabular}
\end{table}

\bigskip

The electrostatic models may thus be defined with focus on the laws of Coulomb
and Biot-Savart (like in the previous subsection) or alternatively as
approximations of Maxwell's equations. These definitions are summarized in
Table~\ref{tab:table1}. The proof that the A and B columns define the same
models involves only standard \ procedures found in basic textbooks. In
electrostatics one start from the Coulomb law and obtain the divergence and
curl of the electric field. By the Helmholtz theorem we have the equivalence
(assuming appropriate conditions at infinity \cite{Infinity})
\begin{equation}
(\ref{coul})\Longleftrightarrow(\ref{gaussE})\text{ and }(\ref{rotE})
\label{impl1}%
\end{equation}
In the same way we of course have $(\ref{EC})\Leftrightarrow(\ref{ECdiff}).$
In magnetostatics one starts from the law of Biot-Savart and obtain the
divergence and curl of the magnetic field. One finds the equivalence
\begin{equation}
(\ref{biot})\text{ and }(\ref{contstat})\Longleftrightarrow(\ref{gaussB}%
)\text{ and }(\ref{amp}) \label{impl3}%
\end{equation}
The above relations trivially remain valid if we allow for time-dependence
where time appears only as a parameter. However, in a time-dependent situation
it is logical to use the general continuity equation (\ref{cont}) instead of
the static one (\ref{contstat}). Then instead of (\ref{impl3}) we find%
\begin{equation}
(\ref{biot})\text{ and }(\ref{cont})\Longleftrightarrow(\ref{gaussB})\text{
and }(\ref{ampdarw}) \label{impl4}%
\end{equation}
Thus the Amp\'{e}re-Darwin equation is a mathematical implication of Biot-Savart's law combined with the continuity equation. The relation (\ref{impl4}) is not included in most standard textbooks but
has (more or less explicitly) appeared in several contributions to this Journal.\cite{Mello,Biswas,Gauthier}.

The equation (\ref{EF}) for the Faraday electric field is formally analogous
to the Biot-Savart law (\ref{biot}) for the magnetic field. Formally similar
to (\ref{impl3}) is the equivalence
\begin{equation}
(\ref{EF})\text{ and }(\ref{gaussB})\iff(\ref{EFdiff})\text{ and
}(\ref{gaussB}) \label{impl5}%
\end{equation}

Table~\ref{tab:table2} includes some laws of electromagnetics and states if
they are valid within EQS, MQS or Darwin. All the included equations are
familiar from standard electromagnetics with the exception of 
Amp\'{e}re-Darwin's equation (\ref{ampdarw}).

\begin{table}[ptb]
\caption{The table includes several familiar laws from electromagnetics as
well as the Maxwell-Darwin equation and state if they are valid within the
EQS, MQS and Darwin models.}%
\label{tab:table2}%
\begin{ruledtabular}
\begin{tabular}{ccccc}
\textbf{Eq.\#} & \textbf{Equation} & \textbf{EQS} & \textbf{MQS} &
\textbf{Darwin}\\
\hline
(\ref{coul}) & Coulomb's law & yes & no & no\\
(\ref{biot}) & Biot-Savart's law & yes & yes & yes\\
(\ref{cont}) & Continuity eq. & yes & no & yes\\
(\ref{contstat}) & $\nabla\cdot\mathbf{J=0}$ & no & yes & no\\
(\ref{gaussE}) & Gauss law & yes & yes & yes\\
(\ref{gaussB}) & $\nabla\cdot\mathbf{B=0}$ & yes & yes & yes\\
(\ref{faraday}) & Faraday's law & no & yes & yes\\
(\ref{ampmax}) & Amp\'{e}re-Maxwell & yes & no & no\\
(\ref{rotE}) & $\nabla\times\mathbf{E=0}$ & yes & no & no\\
(\ref{amp}) & Amp\'{e}re's law & no & yes & no\\
(\ref{ampdarw}) & Amp\'{e}re-Darwin & yes & no & yes\\
\end{tabular}
\end{ruledtabular}
\end{table}

\subsection{A limit of instantaneous propagation in Maxwell's equations}

In subsection A above we obtained the Darwin model from EQS by including
Faraday's law in a simple and straightforward way. An alternative procedure is
to follow the standard textbook derivation of Maxwell's
equations\cite{Griffiths} but in the final step, when the displacement current
is added to Amp\'{e}re's law for consistency with the continuity equation, find
the Amp\'{e}re-Darwin equation (\ref{ampdarw}) rather than the Amp\'{e}re-Maxwell
equation (\ref{ampmax}). This replaces the usual \textit{maximal} assumption
in the correcting term\cite{Zapolsky} with a kind of \textit{minimal}
assumption. Both these derivations of Darwin's model may however seem rather
superficial and arbitrary.\ It would be nice to obtain it from Maxwell's
equations by using some more systematic method. A basic feature of quasistatic
approximations to Maxwell's equations is the instantaneous propagation of
fields. Thus it should be possible to consider quasistatics as some limit
$c\rightarrow\infty$ of Maxwell's equations. This is however quite a singular
limit and the procedure must be further specified. We take the absence of
time-retardation as being the most salient property of quasistatics. Let us express Maxwell's
equations in terms of the potentials $\left(  V,\mathbf{A}\right)  $ so that
the retarded time appears explicitly.\ We choose to use the Coulomb gauge%

\begin{equation}
\nabla\cdot\mathbf{A}=0\label{coulgauge}%
\end{equation}
and follow the derivation in Nielson and Lewis\cite{NielsonLewis} (the reason
for \textit{not} using the Lorentz gauge is considered soon). The equations
for the potentials become (see Griffiths\cite{Griffiths} p. 421),%
\begin{equation}
\nabla^{2}V=-\frac{\rho}{\varepsilon_{0}}\label{poisson}%
\end{equation}%
\begin{equation}
\nabla^{2}\mathbf{A}-\mu_{0}\varepsilon_{0}\frac{\partial^{2}\mathbf{A}%
}{\partial t^{2}}=-\mu_{0}\mathbf{J}+\mu_{0}\varepsilon_{0}\nabla
\frac{\partial V}{\partial t}\label{Awave}%
\end{equation}
Time-retardation appears explicitly if we solve equation (\ref{Awave}) for
the vector potential in terms of an integral in the usual way. The omission of
retarded time in this integral is the same as excluding the second order
time-derivative so that%

\begin{equation}
\nabla^{2}\mathbf{A}=-\mu_{0}\mathbf{J}+\mu_{0}\varepsilon_{0}\nabla
\frac{\partial V}{\partial t} \label{Adarwin}%
\end{equation}
The model so obtained consists of the equations (\ref{coulgauge}),
(\ref{poisson}) and (\ref{Adarwin}). These constitute the
Darwin model in terms of potentials.\cite{error} The usual Darwin equations
(\ref{gaussE}), (\ref{gaussB}), (\ref{faraday}) and (\ref{ampdarw}) is
obtained by use of%
\begin{equation}
\mathbf{E}=-\nabla V-\frac{\partial\mathbf{A}}{\partial t}\text{,
\ \ }\mathbf{B}=\nabla\times\mathbf{A}\text{, \ \ }\mathbf{E}_{C}=-\nabla V
\label{EBEC}%
\end{equation}

Why didn't we use the Lorentz gauge? It is straightforward to make the
corresponding calculations also in that case. We then get a model
approximating Maxwell's equations which seems pretty close to the Darwin model
with the Amp\'{e}re-Darwin law as one of the equations. However, Gauss law
(\ref{gaussE}) is not obtained but a new term appears in the
corresponding equation. Certainly we prefer not to change Gauss law and this
problem, as we have seen, does not appear if we use the Coulomb gauge in the
approximation procedure. Using the Coulomb gauge in the Darwin model also have
the nice consequence that the Coulomb part of the E-field is given by the
scalar potential $\mathbf{E}_{C}=-\nabla V$ and the Faraday part by the vector
potential $\mathbf{E}_{F}=-\partial\mathbf{A}/\partial t.$

\subsection{The potential representations of EQS, MQS and Darwin}

We use the Coulomb gauge. The magnetic field is written $\mathbf{B}%
=\nabla\times\mathbf{A}$ for all models. For MQS the vector potential then
satisfy%
\begin{equation}
\nabla^{2}\mathbf{A}=-\mu_{0}\mathbf{J} \label{AEQSMQS}%
\end{equation}
or in integral form%
\begin{equation}
\mathbf{A}=\dfrac{\mu_{0}}{4\pi}\iiint\dfrac{\mathbf{J}}{R}d\tau^{\prime}
\label{AEQSMQSint}%
\end{equation}
For EQS and Darwin the corresponding equations are (\ref{Adarwin}) or 
\begin{equation}
\mathbf{A}=\dfrac{\mu_{0}}{4\pi}\iiint\dfrac{(\mathbf{J}+\mathbf{J}%
\cdot\mathbf{\hat{R}\hat{R}})}{2R}d\tau^{\prime} \label{AdarwinInt}%
\end{equation}
It takes some manipulation involving partial integration and the continuity
equation to derive (\ref{AdarwinInt}) from (\ref{Adarwin}). The scalar
potential is just the Coulomb potential for EQS, MQS and Darwin,%
\begin{equation}
V=\dfrac{1}{4\pi\varepsilon_{0}}\iiint\dfrac{\rho}{R}d\tau^{\prime} \label{VC}%
\end{equation}
The electric field for EQS is conservative and is just the Coulomb field%
\begin{equation}
\mathbf{E}=-\nabla V \label{Ecoul}%
\end{equation}
while the electric field for MQS and Darwin also include the magnetically
induced electric field so that%
\begin{equation}
\mathbf{E}=-\nabla V-\frac{\partial\mathbf{A}}{\partial t}
\label{Ecoulfaraday}%
\end{equation}

\subsection{Fields of a moving point charge}

We will now consider the quasistatic fields from a point
charge in general motion. Coulomb's law is often taken as a starting point for electrostatics. It is conceptually simple to build on interactions between point charges in this way, so
why not develop all of electrodynamics by generalizing this approach? This is
discussed by Griffiths (chapter 2).\cite{Griffiths} Two major problems are

\begin{itemize}
\item[1.] The force between two point charges depends not only on their
separation but also on both their velocities and accelerations

\item[2.] Furthermore, it is the position, velocity\ and acceleration at
\textit{retarded time} of the other particle that matters.
\end{itemize}

The second point is far more challenging than the first one. However,
time-retardation vanishes within quasistatics and the approach with
interacting point charges become quite simple and instructive. In the point
charge approach to electrodynamic models we automatically include the
continuity equation (\ref{cont}). Therefore this section concerns EQS and
Darwin but not MQS. Let us start with the EQS model. The fields from a point
charge $Q$ within EQS is%
\begin{equation}
\mathbf{E}\left(  \mathbf{r},t\right)  =\dfrac{Q}{4\pi\varepsilon_{0}}%
\dfrac{\mathbf{\hat{R}}}{R^{2}} \label{coulQ}%
\end{equation}%
\begin{equation}
\mathbf{B}\left(  \mathbf{r},t\right)  =\dfrac{\mu_{0}Q}{4\pi}\dfrac
{\mathbf{v}\times\mathbf{\hat{R}}}{R^{2}} \label{biotQ}%
\end{equation}
where the notations in (\ref{Rdef}) are used. The position of $Q$ is
$\mathbf{r^{\prime}}$ which is a function of time and $\mathbf{v}%
=d\mathbf{r^{\prime}}/dt$. From these equations we obtain (\ref{coul}) and (\ref{biot})
while the continuity equation (\ref{cont}) is implied by the point charge description.

Let us now consider the Darwin model. Then equation (\ref{biotQ}) remains
valid while (\ref{coulQ}) only gives the Coulomb part of the electric field%
\begin{equation}
\mathbf{E}_{C}\left(  \mathbf{r},t\right)  =\dfrac{Q}{4\pi\varepsilon_{0}%
}\dfrac{\mathbf{\hat{R}}}{R^{2}} \label{coulQC}%
\end{equation}
The magnetically induced part $\mathbf{E}_{F}$ must now
also be included. However, it is then convenient to start all over again using
the electromagnetic potentials. The potentials are
\begin{equation}
V\left(  \mathbf{r},t\right)  =\dfrac{Q}{4\pi\varepsilon_{0}R} \label{potVQ}%
\end{equation}%
\begin{equation}
\mathbf{A}\left(  \mathbf{r},t\right)  =\dfrac{\mu_{0}Q}{4\pi}\dfrac
{\mathbf{v+v}\cdot\mathbf{\hat{R}\hat{R}}}{2R} \label{potAQ}%
\end{equation}
The scalar potential (\ref{potVQ}) is just the Coulomb potential and the
vector potential (\ref{potAQ}) may be found from (\ref{AdarwinInt}). An alternative derivation is by the ansatz
\begin{equation}
\mathbf{A}\left(  \mathbf{r},t\right)  =\dfrac{\mu_{0}Q}{4\pi}\dfrac
{\mathbf{v}}{R}+\nabla\phi\label{potAQ1}%
\end{equation}
where the scalar field $\phi$ is determined by the Coulomb gauge condition
(\ref{coulgauge}). It takes a little algebra to find
\[
\phi=-\dfrac{\mu_{0}Q}{8\pi}\mathbf{v}\cdot\mathbf{\hat{R}}%
\]
and thus (\ref{potAQ}).\cite{Breitenberger} The electromagnetic fields is related to the potentials in the usual way (\ref{EBEC}). The magnetically induced part of the
electric field is$\ $%
\begin{equation}
\mathbf{E}_{F}=-\partial\mathbf{A}/\partial t \label{EFQ}%
\end{equation}
This expression will obviously involve the acceleration $\mathbf{a}=d\mathbf{v}/dt$ of
$Q$. The Darwin model should thus follow from the potentials (\ref{potVQ}) and
(\ref{potAQ}) of a point charge and it is sufficient to check (\ref{biot}%
)-(\ref{EF}). Only equation (\ref{EF}) is not obvious and the easiest approach
is to use the equivalent differential equations (\ref{EFdiff}). But these
equations are trivially satisfied by (\ref{EFQ}) using the Coulomb gauge and
$\mathbf{B=\nabla\times A.}$

The standard textbook approach to electrostatics begins with Coulomb's law (\ref{coulQ}) for a charge at rest. Sometimes the corresponding approach to magnetostatics is used by taking (\ref{biotQ}) as a starting point but now, of course, with a moving charge. This
may in principle seem wrong since this is a
time-dependent system. Most textbooks take this seriously and starts instead from the
Biot-Savart law with stationary current. However, leaving the point
particle point of view also makes magnetostatics more difficult than
electrostatics. It comes as a relief that the magnetostatic differential
equations (\ref{gaussB}) and (\ref{amp}) are formally similar to, and not much
more difficult than, the electrostatic equations (\ref{gaussE}) and
(\ref{rotE}). Quasistatics is easier than statics in this respect; it allows
for the use of (\ref{biotQ}).

A conventional Lagrangian description of the electromagnetic interaction
between two or more charged particles is possible only if time-retardation may
be neglected. It is straightforward to give such a formulation constructed
from the potentials (\ref{potVQ}) and (\ref{potAQ}).\cite{Kaufman} This results in the so
called Darwin Lagrangian that was first obtained by Oliver Heaviside
in 1891.\cite{Jackson,Heaviside} It was found again by C.G. Darwin in 1920 by
an expansion of the Lienard-Wiechert potentials.\cite{Darwin} \ Rather
surprisingly it then turns out that that terms up to the order $(v/c)^{2}$
inclusive are to be kept in the Darwin Lagrangian (see Jackson\cite{Jacksons
bok} p.596).

\subsection{The Poynting theorem}

Let us follow the standard derivation of Poynting's theorem while using the
quasistatic models. I.e. we add the two equations obtained by taking the
scalar product of the curl E equation with the B-field and the scalar product
of the curl B equation with the E-field. We the get (after the usual
manipulations) for EQS%
\begin{equation}
\frac{\partial}{\partial t}(\frac{1}{2}\varepsilon_{0}E^{2})+\nabla\cdot
(\frac{1}{\mu_{0}}\mathbf{E}\times\mathbf{B})=-\mathbf{E}\cdot\mathbf{J}
\label{EQSenergy}%
\end{equation}
for MQS we obtain%
\begin{equation}
\frac{\partial}{\partial t}(\frac{1}{2\mu_{0}}B^{2})+\nabla\cdot(\frac{1}%
{\mu_{0}}\mathbf{E}\times\mathbf{B})=-\mathbf{E}\cdot\mathbf{J}
\label{MQSenergy}%
\end{equation}
and for the Darwin model%
\begin{equation}
\frac{\partial}{\partial t}(\frac{1}{2}\varepsilon_{0}E_{C}^{2}+\frac{1}%
{2\mu_{0}}B^{2})+\nabla\cdot(\frac{1}{\mu_{0}}\mathbf{E}\times
\mathbf{B+\varepsilon}_{0}\frac{\partial V}{\partial t}\frac{\partial
\mathbf{A}}{\partial t})=-\mathbf{E}\cdot\mathbf{J} \label{Darwinenergy}%
\end{equation}

In EQS only the electric field is associated with energy. Building up a
magnetic field costs no energy in this model due to the absence of a
counteracting induced E-field. In MQS only the magnetic field has energy.
Changes of the electric field is associated with changes in the charge
density. However, the continuity equation is not satisfied and there is no
charge transport for which we could calculate the required energy.\ Finally
the Darwin model includes both magnetic and electric energy. Note that only
the Coulomb-part of the E-field is associated with energy. An important
qualitative difference between EQS and MQS on the one side and Darwin's model
on the other is the possibility of natural resonances in the latter. Since
Darwin includes both capacitive and inductive features we may in principle use
these equations to model, for example, some field theoretic manifestation of a
LC circuit.

\section{If Maxwell had worked in between Amp\'{e}re and Faraday}

We consider in this section various procedures to find the full set of Maxwell
equations starting from the laws of Coulomb and Biot-Savart. The
quasistatic perspective is useful and the models EQS, MQS and Darwin will
appear as intermediate stages. The title of this section refers to the
possibility of introducing the displacement current before Farday's
law.\cite{JammerStachel}

Let us start with the standard textbook procedure of finding Maxwell's
equations.\cite{Griffiths} \ From the static laws of Coulomb and Biot-Savart
we find, by use of the static continuity equation, the static limit of
Maxwell's equations. By introducing magnetic induction we then obtain MQS
(i.e. "Electrodynamics before Maxwell" in Griffiths textbook). The final step,
motivated by the need for consistency with charge conservation (\ref{cont}),
is to introduce the displacement current. At this point a question mentioned
already in the introduction may surface: Maxwell's equations includes waves in
\textit{free space} but why should these follow from \textit{charge
conservation}? Quasistatics is useful for discussing this issue. It is indeed
not necessary to introduce all of the displacement current to save charge
conservation. It is sufficient to include the Coulomb part of
it\cite{Zapolsky} and then we get the Darwin model. At this point no
surprising new physics appears. The step from Darwin to the Maxwell's
equations may then, to begin with, be motivated by symmetry and beauty of
equations. The explicit appearance of the Coulomb electric field in the Darwin
model is a rather unsatisfactory feature and it is so easily fixed by just
replacing it with the total electric field. Amazing new physics now appears
and \ this is accompanied with beautiful new mathematical structure like the
Lorentz invariance of both the Maxwell equations and of the trajectories of
test charges. This is of course surprising but, at least, the previous rather mystifying
motivation for it has now been removed.

Jammer and Stachel discuss what might have happened if Maxwell had worked in
between Amp\'{e}re and Faraday.\cite{JammerStachel} As in the standard
textbook derivation of Maxwell's equations one may first derive the static
limit of Maxwell's equations from the laws of Coulomb and Biot-Savart combined
with the static continuity equation. At this stage Maxwell might have added
the displacement current before the discovery of Faraday's law. This would
result in the EQS model which, according to Jammer and Stachel, is exactly
Galilei invariant. They suggest in the article abstract that the discovery of
Faraday "would have confronted physicists with the dilemma: give up the
Galilean relativity principle for electromagnetism (ether hypothesis), or
modify it (special relativity). This suggest a new pedagogical approach to
electromagnetic theory, in which the displacement current and the Galilean
relativity principle are introduced before the induction term is discussed."
This approach is however less striking than it first seems. We like to include
the appropriate invariance structure not only for the field equations but also
for the trajectories of test charges. Here one get problems with the Galilei
invariance of EQS while the Lorentz invariance of Maxwell's equations is
perfect. The problem of finding satisfactory quasistatic and Galilei invariant
approximations to Maxwell's equations is not an easy one.\cite{Le Bellac}

A third approach to Maxwell equations is of more quasistatic nature. We avoid
in this derivation the static limit of Maxwell's equation by allowing for
(trivial) time-dependence\ in the laws of Coulomb (\ref{coul}) and Biot-Savart
(\ref{biot}). We obtain the differential equations of EQS (i.e. the equations
(\ref{gaussE}), (\ref{gaussB}), (\ref{ampmax})) and (\ref{rotE}) by pure
mathematics if we also use the (time-dependent) continuity
equation\ (\ref{cont}). Maxwell's equations now follow directly when the
Faraday law is introduced. Also in this case (like in the standard textbook
derivation) a question may appear. Why do the magnetic induction experiments
of Faraday imply electromagnetic waves and time-retardation? The answer is
somewhat more hidden in this case. The electric field in EQS is a pure Coulomb
field $\mathbf{E}_{C\text{ }}$but when we include magnetic induction we also get
the Faraday electric field $\mathbf{E}_{F}$. It is now not clear if we should have $\mathbf{E}$ or $\mathbf{E}_{C\text{ }}$ in
(\ref{ampmax}). In the latter case (the \textit{minimal} assumption) we get
the Darwin model and no qualitatively new physics appears (only what is needed
to explain the experiments of Faraday). The step from Amp\'{e}re-Darwin to the
Amp\'{e}re-Maxwell law has already been discussed.

\section{The laws of Coulomb and Biot-Savart in time-dependent theory}

The laws of Coulomb and Biot-Savart provide a starting point for the static
theory of electromagnetics. Let us consider the following two questions:

\begin{itemize}
\item[1.] How should the static laws of Coulomb and Biot -Savart be
generalized to time-dependent theory?

\item[2.] Consider the trivial generalization given by the equations
(\ref{coul}) and (\ref{biot}) where time is included as a parameter. What is
the significance of these formulas?
\end{itemize}

These questions are usually not addressed in the textbooks. The first question
is considered by Jefimenko and the resulting formulas involves (as must be
expected) integrals where the retarded time appears.\cite{Jefimenko}
Jefimenko's generalized laws of Coulomb and Biot-Savart are used by Griffiths
and Heald\ to address the second question. They find that\cite{GriffithsHeald}

\begin{enumerate}
\item[(a)] The generalized Biot-Savart law reduces to the standard one if
$\partial^{2}\mathbf{J}$/$\partial t^{2}=0$ (then, from the continuity
equation, also $\partial^{3}\rho/\partial t^{3}=0$).

\item[(b)] The generalized Coulomb law reduces to the
standard one if $\ \partial\mathbf{J}$/$\partial t=0$ (and thus also
$\partial^{2}\rho/\partial t^{2}=0$).

\item[(c)] The law of Amp\'{e}re holds if $\partial^{2}\mathbf{J}$/$\partial t^{2}=0$ and $\partial
\rho/\partial t=0$.
\end{enumerate}

These results are all sufficient conditions for the laws but only (c) is also necessary. Below
necessary and sufficient conditions are formulated. It is natural to consider
the validity of Coulomb and Biot-Savart in time-dependent theory from the
perspective of quasistatics. We may divide electromagnetics into three
regimes: statics, quasistatics and high frequency. However, statics is just a
particular case of the general Maxwell equations while quasistatics is an
approximation. This motivate a third question :

\begin{enumerate}
\item[3.] When does it happen that a solution in quasistatics is an exact
solution to Maxwell's equations?
\end{enumerate}

It will be shown that a solution of the Darwin model also solves Maxwell's
equations if and only if the current has the form%

\begin{equation}
\mathbf{J}\left(  \mathbf{r},t\right)  =\mathbf{a\left(  \mathbf{r}\right)
}t+\mathbf{b\left(  \mathbf{r}\right)  }-\varepsilon_{0}\frac{\partial
\mathbf{E}_{C}\left(  \mathbf{r},t\right)  }{\partial t}\label{J1}%
\end{equation}
where the Coulomb field is defined by (\ref{EC}) and the vectorfields
$\mathbf{a}$ and $\mathbf{b}$ satisfy $\nabla\cdot\mathbf{a}=\nabla
\cdot\mathbf{b}=0$. Formally we write this%
\begin{equation}
\text{Maxwell + Biot-Savart}\iff\text{ Darwin + Eq.(\ref{J1})}\label{J1Equiv}%
\end{equation}
Let us compare this with (a) above. The last term in (\ref{J1}) may
appear new and unfamiliar in the context of being a part of the true current
(this term but with opposite sign is familiar as a part of the displacement
current). However, this term is just the solenoidal part of the current. Thus
consider any current density $\mathbf{J}$ and write it uniquely in accordance
with Helmholtz theorem as the sum of two parts $\mathbf{J}=\mathbf{J}%
_{T}+\mathbf{J}_{L}$ where $\mathbf{J}_{T}$ is irrotational and $\mathbf{J}%
_{L}$ is solenoidal.\cite{Infinity,dEC/dt} Then it follows from the continuity
equation and the time-derivative of (\ref{ECdiff}) that $\mathbf{J}%
_{L}=-\varepsilon_{0}\partial\mathbf{E}_{C}/\partial t$. The result (a) may be
reformulated in a way similar to (\ref{J1Equiv}) but then we only get the left
implication (i.e."$\Leftarrow$") and the last term in (\ref{J1}) is then assumed to be linear in time. The leaking capacitor is one of a few examples below
showing that quasistatics may apply exactly also with a nonlinear
time-dependence of the current.

Let us now prove (\ref{J1Equiv}) and start with the right implication. That
Darwin is satisfied follows directly from the equivalence (\ref{impl4}). Then
both Amp\'{e}re-Maxwell and Amp\'{e}re-Darwin are satisfied implying $\partial
\mathbf{E}_{F}\mathbf{/}\partial t=0$. \ From (\ref{EFdiff}) it then follows
that $\partial^{2}\mathbf{B/}\partial t^{2}=0$. From the second order
time-derivative of Amp\'{e}re-Darwin and (\ref{gaussB}) we now get%
\begin{equation}
\frac{\partial^{2}}{\partial t^{2}}\left(  \mathbf{J}+\varepsilon_{0}%
\frac{\partial\mathbf{E}_{C}}{\partial t}\right)  =0\label{J1dt2}%
\end{equation}
and (\ref{J1}) easily follows (including the conditions $\nabla\cdot
\mathbf{a}=\nabla\cdot\mathbf{b}=0)$. Consider next the left implication in
(\ref{J1Equiv}). By substituting (\ref{J1}) in the Maxwell-Darwin equation we
find $\nabla\times\mathbf{B}=\mu_{0}(\mathbf{a}t+\mathbf{b})$ so that
$\mathbf{B}$ is at most linear in $t$. From the time-derivative of
(\ref{EFdiff}) we then get $\partial\mathbf{E}_{F}\mathbf{/}\partial t=0$. In
this case the Amp\'{e}re-Darwin and the Amp\'{e}re-Maxwell laws are the same and the
solution to Darwin also solves Maxwell's equations.

Above we found a necessary and sufficient condition (\ref{J1Equiv}) for the exact validity of Biot-Savart's law within Maxwell's equations. Thereby the result (a) of Griffiths and Heald was generalized. Let us now consider the corresponding result for Coulomb's law and allow for a current of the form%
\begin{equation}
\mathbf{J}\left(  \mathbf{r},t\right)  =\mathbf{b\left(  \mathbf{r}\right)
}-\varepsilon_{0}\frac{\partial\mathbf{E}_{C}\left(  \mathbf{r},t\right)
}{\partial t}\label{J2}%
\end{equation}
where $\nabla\cdot\mathbf{b}=0$. We will prove that%
\begin{equation}
\text{Maxwell + Coulomb }\iff\text{ EQS + Eq.(\ref{J2})}\label{J2Equiv}%
\end{equation}
Let us start with the right implication in \ (\ref{J2Equiv}). From Coulomb's
law it follows the E-field is conservative so that EQS is obtained from
Maxwell. From Faraday's law it then follows that $\partial\mathbf{B/}\partial
t=0$ and from the time-derivative of Amp\'{e}re-Maxwell we now get (using
$\mathbf{E}=\mathbf{E}_{C}$)%
\begin{equation}
\frac{\partial}{\partial t}\left(  \mathbf{J}+\varepsilon_{0}\frac
{\partial\mathbf{E}_{C}}{\partial t}\right)  =0\label{J2dt}%
\end{equation}
and (\ref{J2}) easily follows. To obtain the left implication of
(\ref{J2Equiv}) we first substitute (\ref{J2}) in Amp\'{e}re-Maxwell and obtain.
$\nabla\times\mathbf{B}=\mu_{0}\mathbf{b}$. This equation together with (\ref{gaussB}) results in a time-independent
B-field and thus Faraday's law is satisfied. This is what is needed for the
Maxwell equations to be valid for an EQS solution. The proof of (\ref{J2Equiv}) is now completed and the result (b) is generalized.

Let us finally consider the result (c) involving
Amp\'{e}re's law. This condition turns out to be not
only sufficient but also necessary. Consider a current density of the form%
\begin{equation}
\mathbf{J}\left(  \mathbf{r},t\right)  =\mathbf{a\left(  \mathbf{r}\right)
}t+\mathbf{b\left(  \mathbf{r}\right)  } \label{J3}%
\end{equation}
where the vectorfields $\mathbf{a}$ and $\mathbf{b}$ satisfy $\nabla
\cdot\mathbf{a}=\nabla\cdot\mathbf{b}=0$. Then%
\begin{equation}
\text{Maxwell + Amp\'{e}re }\iff\text{MQS + Eq.(\ref{J3}) + Eq.(\ref{cont})}
\label{J3Equiv}%
\end{equation}
Consider first the right implication. Both Amp\'{e}re's law and Amp\'{e}re-Maxwell are
valid so $\partial\mathbf{E}/\partial t=0$. By using the time-derivative of
Faraday's law we then find $\partial^{2}\mathbf{B/}\partial t^{2}=0$ so from
the second order time-derivated Amp\'{e}re's law we get $\partial^{2}%
\mathbf{J}/\partial t^{2}=0.$ Then (\ref{J3}) follows if we also use that the
current has zero divergence. The left implication in (\ref{J2Equiv}) is
obtained by just reversing the procedure above.

We will now consider a few examples involving the results above. The first one will also be used in the next subsection.

\begin{example}
Consider an (infinitely) long solenoid or a toroidal coil with time-varying current
$I\left(  t\right)  $. We assume the coils are winded so that we may neglect
the axial current in the solenoid and the poloidal current in the toroid.
\end{example}

Within quasistatic we get (exactly) zero B-field outside the coils (from the Biot-Savart
law and symmetry). The MQS and Darwin models coincide because there is no charge density and accordingly $\mathbf{E}_{C}=0$. There is a
magnetically induced electric field $\mathbf{E}=\mathbf{E}_{F}$ outside
the coils in spite the fact that the B-field vanishes exactly.
In general this is only an approximative solution to the Maxwell equations
since the Amp\'{e}re law (and of course the Amp\'{e}re-Darwin law) rather than the
Amp\'{e}re-Maxwell law is satisfied outside the coils. An exact solution is obtained, as we have seen, when
the current is linear in time. The electric field is then constant in time.

The second example shows that quasistatics may agree exactly
with Maxwell's equations even when he time-variation is quite arbirary.
\begin{example}
Consider any given spherically symmetric charge and current density satisfying
the continuity equation (\ref{cont}).
\end{example}

By the spherical symmetry we get $\nabla\times\mathbf{E}=\nabla
\times\mathbf{B}=0$. The solution to EQS, Darwin and Maxwell's equations is
$\mathbf{E}=\mathbf{E}_{C}$, $\mathbf{B}=0$ and $\mathbf{J}=-\varepsilon
_{0}\partial\mathbf{E}_{C}/\partial t$ which is valid for any prescribed
charge density $\rho=\rho\left(  r,t\right)  $ in (\ref{EC}). The equivalences
(\ref{J1Equiv}) and (\ref{J2Equiv}) apply with $\mathbf{a}%
=\mathbf{b}=0$. A particular case is a leaking spherical capacitor. We assume that the medium
in between the spherical plates has some conductance and does not violate the spherical symmetry. In
between the plates there is, during the slow discharge, zero charge density.
The Amp\'{e}re law is clearly not satisfied because $\mathbf{B}=0$ while
$\mathbf{J}\neq0$.

The examples 3-5 below illustrates that in a linear, isotropic and homogeneous conductor
 the conduction current (assuming Ohm's law), in certain cases, does not create any
magnetic field. This corresponds to exact solutions of Maxwell's equations that also
solves EQS and Darwin.

\begin{example}
Let us now consider an example without symmetry in the charge distribution.
Space is assumed to be linear, isotropic, conducting and homogeneous
characterized by $\left(  \varepsilon_{0},\mu_{0},\sigma\right)  $. Assume
that we at time $t=0$ know the charge density $\rho\left(  \mathbf{r}%
,0\right)  =\rho_{0}\left(  \mathbf{r}\right)  $.
\end{example}

From $\mathbf{J}=\sigma\mathbf{E}$ in the continuity equation and by use of
Gauss law we find%
\begin{equation}
\rho\left(  \mathbf{r},t\right)  =e^{-\left(  \sigma/\varepsilon_{0}\right)
t}\rho_{0}\left(  \mathbf{r}\right)
\end{equation}
The corresponding solution to Maxwell's equation is%
\begin{equation}
\mathbf{E}\left(  \mathbf{r},t\right)  =e^{-\left(  \sigma/\varepsilon
_{0}\right)  t}\mathbf{E}_{0}\left(  \mathbf{r}\right)  \text{, \ \ \ \ \ \ }%
\mathbf{B}=0
\end{equation}
where $\mathbf{E}_{0}\left(  \mathbf{r}\right)  $ is the Coulomb field due to
the charge density $\rho_{0}\left(  \mathbf{r}\right)  $.

\begin{example}
The last example above may be modified so that we are dealing with a finite
conductor surrounded by free space. Let the conductor contain all given charge
$\rho_{0}\left(  \mathbf{r}\right)  $ at time $t=0$ and let the surface of the
conductor be an equipotential surface of $V_{0}\left(  \mathbf{r}\right)  $
defined by%
\begin{equation}
V_{0}\left(  \mathbf{r}\right)  =\dfrac{1}{4\pi\varepsilon_{0}}\iiint
\dfrac{\rho_{0}\left(  \mathbf{r^{\prime}}\right)  }{R}d\tau^{\prime}%
\end{equation}

\end{example}

Inside the conductor both the charge distribution and the solution to
Maxwell's equations is the same as in the previous example. Outside the fields
are static with $\mathbf{E}=\mathbf{E}_{0}\left(  \mathbf{r}\right)  =-\nabla
V_{0}$ and
$\mathbf{B}=0\mathbf{.}$ The surface charge density on the conductor is%
\begin{equation}
\sigma_{S}\left(  \mathbf{r},t\right)  =\varepsilon_{0}\left(  1-e^{-\left(
\sigma/\varepsilon_{0}\right)  t}\right)  \mathbf{\hat{n}}\left(
\mathbf{r}\right)  \cdot\mathbf{E}_{0}\left(  \mathbf{r}\right)
\end{equation}
where $\mathbf{\hat{n}}$ is the outward normal. 

\begin{example}
Related to the last example is the leaking capacitor of arbitrary geometry and
with\ homogeneous, isotropic and weakly conducting material in between the two
perfect conductors.\cite{Bartlett}
\end{example}

In this case we must define $V_{0}\left(  \mathbf{r}\right) $ by Laplace equation with appropiate boundary conditions on the two perfect conductors. The solution may be then expressed in terms of $\mathbf{E}=\mathbf{E}_{0}\left(  \mathbf{r}\right)  =-\nabla
V_{0}$ as in the two last examples above.

\section{A quasistatic perspective on question \#6}

A time-varying current of a long solenoid causes an induced electric field at
the outside. But how can that be? There is no magnetic field at the outside!
A.P French asked a similar question and it was addressed in several
papers.\cite{French,Mills,Walstad,Templin,Protheroe} The answers included two
essential points

\begin{itemize}
\item The vanishing of the B-field is a static phenomena while in the
nonstatic case there is a small magnetic field outside the solenoid.

\item It is not good to consider the time-varying electromagnetic field as a source
for the induced electric field, the source is rather the time-varying
current.\cite{dB/dt}
\end{itemize}

However, even though both these statements are correct in view of the Maxwell
equations, there may remain some uncomfortable feelings. The way we in
practice calculate the induced electric field from Faraday's law make it
natural to think in terms of the time-varying B-field as a source and, even if
there is a small B-field outside the solenoid, \textit{the main part of this
source is well separated in space from the effect we consider} (i.e. the
induced outside E-field). This apparent action at a distance seems to
contradict the local action quality of Maxwell's equations. But of course, a
careful analysis shows that there is no true contradiction involved.

Let us now consider this problem from a different point of view. If we use a
\textit{quasistatic model} in the calculations it may be favorable to also
\textit{think} in terms of this model and not in terms of the full set of
Maxwell's equations. This is analogous to what we often do in other fields of physics. For
example we use classical mechanics  without worrying about quantum
mechanics, Newtonian kinematics without referring to special relativity or
Newtons law of gravitation without thinking about general relativity. It is in the
solenoid example easy to calculate the magnetic and electric fields using
quasistatics. One gets an exact solution within the quasistatic model and a
good approximation to the true Maxwell solution. This quasistatic solution (of
MQS and Darwin) has no magnetic field at all outside the solenoid. This solution is
fundamentally inconsistent with Maxwell's equations since there is a
time-varying E-field in free space where the magnetic field vanishes exactly.
However, the solution is of course consistent with quasistatics. This is an
indication of the quite different nature of quasistatic models and Maxwell's
equations. The PDE's appearing in quasistatics are often elliptic or parabolic
while the Maxwell equations are hyperbolic. The conceptual problems have their
roots in using \textit{quasistatic} \textit{calculations} but interpreting the
result using \textit{Maxwell's equations.}

\section{Electromagnetic simulations}

\subsection{The PC in basic courses}

Electromagnetic simulations are essential to electrical and electronic product
designs in many industries. A broad range of important applications are within
the quasistatic regime like motors, sensors, power
generators, transformer systems and Micro Electro Mechanical Systems (MEMS).
Industrial and scientific applications may involve extensive calculations and
accordingly the need for much computational power. However, the usual PC has
developed enormously and so has the software for simulations with
increasingly user friendly interfaces. Today the PC is a sufficiently powerful
tool for modeling many electromagnetic phenomena within just seconds of computational time. 
 This makes it potentially useful
as an instrument for teaching basic electrodynamics. In the PC lab students may simulate
various electromagnetic phenomena. Examples within statics or quasistatics include the edge effects in the parallel plate capacitor, distributed currents in conductors of various
shapes, the Hall effect, magneto resistance, the appearance of non vanishing
charge density in inhomogeneous conductors, the electric field outside and
surface charge on a conductor with currents, various objects placed in a given
external static or time-varying electromagnetic fields, eddy currents,
inductive heating, magnetic diffusion, magnetic shielding and more.

The use of electromagnetic simulations as an effective pedagogical tool also presupposes some support from theory. One should in particular focus more on how Maxwell's equations are used to formulate well-posed PDE-problems for various physical situations. Most textboks, at least the basic ones, are not much influenced by the appearance of computers. Their guiding principle is to find analytical solutions in simple and instructive cases. The underlying PDE-problem may not be needed explicitly in some examples like when integral laws are used in very symmetric cases or when tricks like mirror charges or mirror currents work. When the PDEs are actually used, and this is much more the case in the more advanced texts, one finds solutions in terms of integrals and series by comparatively tedious calculations involving, for example, special functions and variable separation.

With a numerical PDE-solver available the situation is somewhat different. The formulation of well posed PDE-problems for various phenomena is now motivated without the intention of finding analytical solutions. This is an easy part of the PDE theory for electromagnetics and it is instructive by providing increased insight into the structure of Maxwell's equations. It is in particular important and sometimes straightforward to find how Maxwell's equations reduce thanks to various symmetries. For example, the independence of one Cartesian coordinate or alternatively an axial symmetry results in a decoupling of Maxwell's equations. A simple inspection of the equations written in component form reveals this structure. Such symmetries explain, for example, why the magnetic field of a toroidal coil only has an azimuthal component or why there are TE and TM modes in planar wave-guides. In the next subsection we illustate this by formulating PDE problems for eddy currents.

\subsection{Equations for eddy currents}

Eddy currents and associated phenomena like inductive heating, magnetic
shielding and magnetic diffusion are in practice very common since the basic
ingredients are just a conductor and a time-varying electromagnetic field. Still eddy
currents are not much discussed in the textbooks. An exception is the one by
Smythe where a whole chapter is devoted to eddy currents and analytical
solutions are given in terms of series and integrals.\cite{Smythe} An
interesting method of images may be used to calculate eddy currents in a thin
conducting sheet. The theory was developed by Maxwell and reformulated in
modern terms by Saslow.\cite{Saslow} However, the analytical theory for eddy
currents is still comparatively difficult. Electromagnetic modeling on a PC
is today an attractive and easy way to include more about eddy currents in
basic courses.

Eddy current problems fall into two classes, steady-state and transient. In
steady state analysis (also called time-harmonic analysis) we simply replace
$\partial/\partial t$ with the factor $j\omega$ and allow for complex valued
fields. Maxwell equations becomes time-independent and much easier to solve.
Physically we may obtain a steady-state condition sufficiently long time after
the start of a time-harmonic source of the fields. The initial transient
behavior is not considered in this analysis. The steady state Maxwell
equations may be further simplified in quasistatic situations, this may be
useful for analytical theory but is not of much interest for electromagnetic
simulations. However, in time-dependent (transient) analysis an initial value
PDE must be solved. Then a quasistatic approximation may imply a major
numerical simplification changing a hyperbolic PDE into an parabolic or
elliptic one.

Let us assume that the time-variation is slow enough so that quasistatic
theory applies. But what quasistatic model should we use in the study of eddy
currents; EQS, MQS or Darwin? Obviously not EQS since magnetic induction
(Faraday's law) is outside the scope of that model. What about MQS? This seems
to be the standard model for eddy currents calculation but it may in fact only
be used in some particular cases.\cite{Jacksons bok,Smythe,Baum} A quite common misunderstanding concerning the quasistatic
approximation may partly explain the popularity of MQS.\cite{error}

Let us now consider situations where the use of MQS may be justified. This
happens in certain symmetric cases when MQS and Darwin are equivalent models.
We consider below two examples of such symmetries. The first one include, as a
special case, the situation when Maxwell's theory for eddy currents in thin
conducting sheets applies.\cite{Saslow}

\begin{example}
Consider a body in which the conductivity $\sigma=\sigma(z)$ only varies in
the z-direction. In the $xy-$ directions the body is homogeneous and of
"infinite" extent. The electromagnetic field is created by the use of an
external current density $\mathbf{J}^{e}=J_{x}^{e}(t,x,y,z)\mathbf{\hat{x}%
}+J_{y}^{e}(t,x,y,z)\mathbf{\hat{y}}$ with vanishing divergence; $\nabla
\cdot\mathbf{J}^{e}=0$.
\end{example}

In this example there will appear no charge density and the Darwin model is
equivalent to MQS. The reason is that electric fields will only appear in
directions of constant conductivity (i.e. they have no $z-$component) and they
cause no pile up of charge. The system may be described in terms of the
potentials $A_{x}$ and $A_{y}$ while $V=A_{z}=0$. From equation (\ref{Adarwin})
with $\mathbf{J}=\mathbf{J}^{e}+\sigma(-\nabla V-\partial\mathbf{A}/\partial
t)$ we get the following equations for $A_{x}$ and $A_{y}$%
\begin{align}
-\mu_{0}\sigma\frac{\partial A_{x}}{\partial t}+\nabla^{2}A_{x}  &  =-\mu
_{0}J_{x}^{e}\label{AxAyEddy}\\
-\mu_{0}\sigma\frac{\partial A_{y}}{\partial t}+\nabla^{2}A_{y}  &  =-\mu
_{0}J_{y}^{e}\nonumber
\end{align}

These 3D equations determine the dynamics and may be used to model physical
systems with the above symmetry.

A similar 2D axi-symmetric case may also be formulated.

\begin{example}
We use cylindrical coordinates $(r,\phi,z)$ and a rotationally symmetric
conducting body with conductivity $\sigma=\sigma(r,z)$. The external current
density is $\mathbf{J}^{e}=J^{e}(t,r,z)\mathbf{\hat{\phi}}$.
\end{example}

Also in this case there will appear no charge density. Some potentials
vanishes, $V=A_{r}=A_{z}=0$, and the system may be described in terms of the
vector potential $\mathbf{A}=A(t,r,z)\mathbf{\hat{\phi}}$. This time we get
the equation for the dynamics as%
\begin{equation}
-\mu_{0}\sigma\frac{\partial A}{\partial t}+\frac{1}{r}\frac{\partial
}{\partial r}(r\frac{\partial A}{\partial r})+\frac{\partial^{2}A}{\partial
z^{2}}=-\mu_{0}J^{e} \label{AphiEddy}%
\end{equation}

The equations may easily be studied by using a PC and some 
PDE-solver.\cite{PDEsolver} For example, a straight copper wire above a copper
plate is considered by Backstrom using equation (\ref{AxAyEddy}) and the solver
FlexPDE.\cite{Backstrom} He also considers a similar example using
(\ref{AphiEddy}) for a circular copper wire above a circular copper plate. The
transient fields appearing when a current is turned on in the wire is studied
(pages 75 and 99 in Backstrom's book). The same models are also easily studied
using Comsols Mulphysics with the Electromagnetic Module. These problems are
numerically simple and, of course, there are many other PDE solvers that may
be used. Such numerical tools make it possible to consider many cases
for which there are no analytical solutions. 

The examples above were carefully designed in order to avoid charge density.
This is also true for all eddy current examples in Saslows' paper (see in
particular Appendix B of that paper).\cite{Saslow} Usually there will however
appear time-varying surface charge on conductors with eddy currents and, in
case of an inhomogeneous conductor, there will also appear charge density
inside the conductor. In order to describe the physics we then need to include
the scalar potential in the analysis and MQS cannot be used \ An interesting
possibility is then to use the Darwin model. Such applications of Darwin's
model have been suggested in connection with eddy currents in the human body
caused by high voltage transmission lines.\cite{Raviart} \ The traditional use
of Darwin's model is not eddy currents but concerns charged particle beams and
plasma simulations.

Let us write equations for the Darwin model in term of potentials. The
conducticity of a possibly inhomogeneous conductor is $\sigma=\sigma
(\mathbf{r})$ and that the time-dependent external current $\mathbf{J}%
^{e}=\mathbf{J}^{e}\left(  t,\mathbf{r}\right)  $ is prescribed The total
current may be written $\mathbf{J}=\mathbf{J}^{e}+\sigma(-\nabla
V-\partial\mathbf{A}/\partial t)$. We use, as always in the Darwin model, the
Coulomb gauge $\nabla\cdot\mathbf{A}=0$. The dynamics of the potentials are
determined by the continuity equation
\begin{equation}
-\varepsilon_{0}\frac{\partial}{\partial t}\nabla^{2}V-\nabla\cdot
(\sigma\nabla V+\sigma\partial\mathbf{A}/\partial t)=-\nabla\cdot
\mathbf{J}^{e} \label{V_Eddy}%
\end{equation}
and the Amp\'{e}re-Darwin equation%
\begin{equation}
-\varepsilon_{0}\mu_{0}\frac{\partial}{\partial t}\nabla V-\mu_{0}%
(\sigma\nabla V+\sigma\partial\mathbf{A}/\partial t)+\nabla^{2}\mathbf{A}%
=-\mu_{0}\mathbf{J}^{e} \label{A_Eddy}%
\end{equation}

These equations have a non-standard appearance by containing mixed time and
spatial derivatives. They may however be solved using Comsol Multiphysics.
Actually there is a considerable freedom for the user to enter non-standard
equations in this PDE-solver. This is achieved by making the equations
available also on the "weak form" level which is the natural form for the
finite element method. The numerical solutions appears, at least
qualitatively, to behave in an expected way for the few examples that we have
considered. The practical usefulness of the Darwin model for eddy current
calculations remains however to be proven.

A qualitatively new feature of Darwin's model (as opposed to MQS or EQS) is the
possibility of resonance. Since the Darwin model includes both capacitive and
inductive phenomena it may in principle be used to model systems where the
energy oscillates between the electric and magnetic fields. Consider, for
example, a field theoretic version of a LC-circuit. A suitable design for such
an application may be a resonator in the form of a short coaxial cable where
one end is short circuited by a metal plate and the other end is still
electrically open but with increased capacitance created by the use of two
close parallel plates, connected to the inner and outer conductor,
respectively. This kind of resonator has applications in connection with
electron beam \ devices at microwave frequencies (example 3.4.1 in the
textbook of Haus and Melcher).\cite{HausMelcher} The resonance frequency of a
LC-circuit is \ $f=1/\sqrt{LC}$ but the frequency must be low enough not to
violate the quasistatic assumption. In the above design of a resonator the
most essential method to achieve this (in the absence of dielectrics or
magnetic materials) is to make the capacitance large and thus take the two
capacitor plates very close to each other. 

\section{Summary}

Maxwell's equations are fundamental for the description of electromagnetic
phenomena and valid for an enormous range of spatial and temporal scales. The
static limit of the theory is well defined and of course much easier. The
electric and magnetic fields may in this limit be given by the laws of Coulomb
and Biot-Savart. However, as soon as there is \textit{any} time-dependence we
should \textit{in principle} use the full set of Maxwell's equations with all
their complexities. Time-retardation is a fundamentally important but also a
complicating feature. Using Maxwell's equations means in analytical theory
that even if the effect is small it will not vanish and this makes the theory
unnecessary complicated. In numerical analysis these effects, however small,
may force us to use smaller timesteps (for numerical stability) and expensive
calculations. It is therefore useful to introduce quasistatic approximations
in Maxwell's equations. The quasistatic models are also useful for a better
understanding both of low frequency electrodynamics and for explaining the
transition from statics to the general high frequency electrodynamics. This have
been discussed in the present paper and below we list a few major points.

\begin{enumerate}
\item[(1)] The quasistatic limit of Maxwell's equations is a kind of
$c\rightarrow\infty$ limit obtained by neglecting time-retardation. The Darwin
model is obtained if we use the Coulomb gauge.

\item[(2)] The Darwin model involves both capacitive and inductive features
but there is no radiation and the interactions are instantaneous. Poynting's
theorem for this model shows that there is both electric and magnetic energy,
but the electric energy only includes the Coulomb part of electric field.

\item[(3)] EQS and MQS may be considered as approximations of the Darwin
model. EQS includes capacitance but not inductance while MQS includes
inductance but not capacitance. Poynting's theorems for these models show that
there are only electric energy in EQS and only magnetic energy in MQS.

\item[(4)] The law of Biot-Savart is valid within EQS, MQS and the Darwin model.

\item[(5)] The law of Coulomb is of general significance for quasistatics (EQS, MQS and Darwin)
as is obvious when we use the formulation in terms of force laws (subsection IIA) or use potentials 
(subsection IID).

\item[(6)] The law of Amp\'{e}re is \textit{not} of general significance within
quasistatics. It is valid only in MQS but not within EQS and Darwin.
\cite{error}

\item[(7)] Galilei invariance of quasistatics is a somewhat delicate issue.
Galilei invariance structures may be defined for EQS and MQS but if we like the
force to have the corresponding invariance then only the electric force is
included in EQS and only the magnetic force in MQS.\cite{Le Bellac}

\item[(8)] Quasistatics has important applications in electromagnetic modeling
of transient phenomena.
\end{enumerate}


\begin{thebibliography}{99}
\bibitem {scriptR}The notations in Ref.\onlinecite{Griffiths} has been the
preferred choice in this paper but here is one exception. Griffiths use a
script lower case "r" instead of "$R"$ but unfortunately I have not
found a nice representation of this symbol on my computer.
\bibitem {FaradayIntRemark}It is important to note that the integral
(\ref{EF}) converges. This is evident from the law of Biot-Savart which
implies that the B-field goes to zero as $1/r^{2}$ when $r\rightarrow\infty.$
\bibitem {Infinity}We  consider physical systems where the charges and
currents do not extend to spatial infinity and may  use as boundary conditions
 that the fields approach zero sufficiently fast far away.
\bibitem {CondEQSMQS}This is never a problem according to Ref.
\onlinecite{HausMelcher} but I have not been able to convince myself that this
is true under all circumstances.
\bibitem {Amperes lag def}Amp\'{e}re's law $\nabla\times\mathbf{H}%
=\mathbf{J}_{f}$ differs from (\ref{amp}) if there is a non-vanishing
polarisation current $\partial\mathbf{P}/\partial t$. However, Amp\'{e}re's
law is strictly speaking a static equation and we prefer to use (\ref{amp}) as
a time-dependent generalization with the same name.
\bibitem {ampere-darwin}Like in Ref. \onlinecite{GriffithsHeald} we use the
names Ampere's law for equation (\ref{amp}) and Ampere-Maxwell for
(\ref{ampmax}). Then it seems natural to use the name Ampere-Darwin for
equation (\ref{ampdarw}).
\bibitem {error}There seems to be a quite widespread misunderstanding that
quasistatics, defined by the omission of time-retardation, includes
Amp\'{e}re's law (\ref{amp}) as a valid equation. This is stated or implied by many
textbooks.\ See for example Jackson (Ref.
\onlinecite{Jacksons bok}, p. 218) or Smythe (Ref. \onlinecite{Smythe}, p.
368). However, this is not correct for an obvious reason. Amp\'{e}re's law
implies stationary current and thus constant charge density. But certainly, a
slowly varying charge density cannot force us to include time-retardation.
\end{thebibliography}
\end{document}